\documentclass[twocolumn,english]{revtex4}
\usepackage[T1]{fontenc}
\usepackage[latin1]{inputenc}
\usepackage{float}
\usepackage{graphicx}

\makeatletter

\providecommand{\tabularnewline}{\\}

\usepackage{float}

\usepackage{graphicx}

\usepackage{babel}
\makeatother
\begin{document}

\title{Evolution of the crystal-field splittings in 
the compounds CeX (X=P, As, Sb, Bi),  \\CeY (Y=S, Se, Te)  
and their alloys CeX$_{1-x}$Y$_{x}$}

\author{P. Roura-Bas~$^{a}$ ,V. Vildosola
$^{a}$~~and A. M. Llois~$^{a,b}$}

\affiliation{$^{a}$ Dpto de F\'{\i}sica, Centro
At\'{o}mico Constituyentes,
Comisi\'{o}n Nacional de Energ\'{\i}a At\'{o}mica}

\affiliation{$^{b}$ Dpto de F\'{\i}sica,
Universidad de Buenos Aires, Buenos
Aires, Argentina}

\begin{abstract}
The crystal-field splittings of the monopnictides and monochalcogenides
of Cerium (CeX and CeY) and their alloys (CeX$_{1-x}$Y$_x$) are calculated by
means of an \emph{ab initio} many-body combined technique.  
The hybridization functions
of the 4f states of Cerium with the conduction band for each material are
obtained from first principles within the local density approximation (LDA)
and are used as input for the Anderson impurity model,
which is solved  within a multi-orbital Non-Crossing Approximation (NCA).
This realistic theoretical approach (LDA-NCA) is able to reproduce the
experimental results
for the crystal-field splittings of the CeX and CeY series in agreement
with previous
theoretical calculations. It is also able to describe the non-linear
evolution of the splittings
in the CeX$_{1-x}$Y$_x$ alloys as a function of x. An analysis of the
values of the crystal-field splittings
in all the compounds  can be   done in depth in this contribution, due to a
detailed knowledge of  the band structure
and crystal environment in combination with  many-body physics.

\end{abstract}
\maketitle

\section{Introduction}
One of the important issues of the last years has been to produce realistic 
theoretical descriptions of  strongly correlated electronic systems. These 
descriptions make a point out of   including  detailed electronic structure 
information as well as  many body dynamical effects on an equal footing. Examples 
of materials for which this kind of treatments becomes essential 
are  Cerium compounds. In these systems, the Cerium atom is very
sensitive to the crystalline and chemical environment and the differences in the
interaction strengths  of  the localized 4f states with the conduction band  can  
give rise to a wide variety of behaviors.
A lot of effort has been devoted to implement realistic calculations 
using different approaches \cite{point-charge-1,Han,dmft-review2}. 

In this work, we apply a mixed \emph{ab initio} many-body technique to 
study the monopnictides CeX (X=P, As, Sb,
Bi), the monochalcogenides
CeY (Y=S, Se, Te) and  CeX$_{1-x}$Y$_{x}$ alloys 
which exhibit interesting and
unusual physical  properties. These compounds crystallize
in the simple rock-salt structure making it
possible to study in depth
the influence of the electronic structure on their properties. 
In particular, the heavier compounds, namely  CeSb, CeBi and 
CeTe \cite{Handbook-Actinides}  present a strong magnetic anisotropy, 
great sensitivity to the application  of pressure and to the dilution of
Cerium by non magnetic ions, as well as to the substitution of a 
pnictogen by another pnictogen or by a
chalcogen\cite{pressure-dilution}. These heavier compounds also
show, in particular,  anomalous small values for the 4f crystal-field
splittings ($\Delta_{CF}$) as compared to the other compounds
within the series.    We are interested here in doing a
thorough analysis of the crystal field splittings as well as of the symmetry 
of the 4f states in these  Ce compounds, for  which is
necessary to disentangle the interactions among  the 4f states and the 
environment.  We focus, thereafter,  on   the study of the crystal field 
splitting within the  CeX and CeY series and also on its evolution 
along  the  CeX$_{1-x}$Y$_{x}$ alloys as a function of $x$. 

The anomalous value of the crystal field splitting of the heavier CeX and 
CeY  compounds has been  previously
treated  and understood by Wills and Cooper \cite{point-charge-1,Cooper-Wills}.
These authors showed that the dominant
contribution to the splittings in these monopnictides  can be
obtained from the point-charge (PC) model which is appropriate for
insulators and ionic systems. But the depression of the splitting
along the series can only be understood if hybridization effects are 
taken into account. The PC model accounts well  for the splittings of
rare earth monopnictides when the rare earth goes from Pr to Tb, 
but it fails to describe Ce systems  because in these compounds
the 4f-band hybridization  with the conduction states cannot 
be neglected. Wills and Cooper considered that the total splitting in 
these systems is the result of two independent contributions: the extrapolated 
value from the PC model considering non-hybridized 4f levels 
and the splittings induced by hybridization. They calculated the 4f
hybridization function out of the conduction band density of states obtained 
using the linear muffin-tin-orbital (LMTO) method within the atomic-sphere
approximation (ASA) for the self consistent potential. This last approximation
does not consider  the  anisotropy of the crystalline environment. Within these
calculations  the  authors treated the 4f state as belonging to the 
core . They obtained the  hybridization contribution to $\Delta_{CF}$ on 
the basis of the Anderson hamiltonian solved to  second order in 
perturbation theory. 

In the present work, we obtain the  $\Delta_{CF}$ 's in a different way, 
we perform \textit{ab initio} calculations using the Full
Potential-LAPW method within the local density approximation (LDA)
\cite{Wien-artice} and  treat the 4f states as part of the valence band. We 
then compute the 4f hybridization function following ref.\cite{Gunnarsson-hib,
Han}. This function contains detailed information on the electronic
structure of each system and is used as input for the Anderson impurity
hamiltonian  which is solved within a multiorbital non-crossing approximation
(NCA). This LDA-NCA approach has already been applied to cubic and
tetragonal Ce based systems to calculate $\Delta_{CF}$'s and to obtain 
trends in the Kondo energy scales, yielding results which are in good 
agreement with the available experimental information \cite{Han, Vero-5}.
We calculate the crystal field splittings for CeX and CeY  and 
compare them with the  previous theoretical results and with experimental data.
We also follow  the evolution of $\Delta_{CF}$ with concentration  for the
CeSb$_{1-x}$Te$_{x}$ alloys, which had not been previously theoretically 
approached, and compare the results with experimental data.  We also analyse
the behavior of CeAs$_{1-x}$Se$_{x}$, which has not been previously tackled
either experimentally nor theoretically.

In section 2 we give a brief description of the LDA-NCA technique.
Section 3 is divided into  three parts: a) results for the $\Delta_{CF}$
splittings of the CeX monopnictides, b) Comparison of  the $\Delta_{CF}$
splittings of  the two series, CeX  and CeY  , and c) study of the
evolution of $\Delta_{CF}$  for the CeX$_{1-x}$Y$_{x}$ alloys. Finally,
in section 4 we discuss and conclude.

\section{Method of calculation}

The LDA-NCA approach is an \emph{ab initio}
many body technique that
solves the Anderson impurity problem using as
input the
hybridization function, $\Gamma(\epsilon)$, of
the conduction band 
with the 4f-state of Ce. The hybridization
function is
calculated from first principles whithin the
Density Functional Theory. In this work, the ab initio calculations
are done using  the full
potential linearized augmented
plane waves method (FP-LAPW), as implemented in
the Wien2k code \cite{Wien}. 

For Ce systems the Anderson impurity
hamiltonian has the form:

\[
H=\sum_{k\sigma}\varepsilon_{k\sigma}c_{k\sigma}^{_{^{\dagger}}}c_{k\sigma}+\sum_{m}\varepsilon_{fm}f_{m}^{\dagger}f_{m}+U\sum_{m>m'}n_{fm}n_{fm'}\]

\begin{equation}
+\sum_{k\sigma,m}(V_{k\sigma,m}f_{m}^{\dagger}c_{k\sigma}+H.C.),\label{eq:hia}\end{equation}
where $V_{k\sigma,m}$ is the hopping matrix
element between the
conduction electron states, ($c_{k\sigma}$), and
the 4f orbitals ($f_{m}$),
$\varepsilon_{fm}$ is the corresponding 
4f-energy level with respect to the Fermi
energy and $U$ the on-site Coulomb repulsion of
the \emph{4f} electrons.
$\Gamma(\varepsilon)$ is proportional to the
product $V_{k\sigma,m}^{*}V_{k\sigma,m'}$
and, as suggested by \emph{Gunnarsson et al.}
\cite{Gunnarsson-hib},
it can be estimated from the projected LDA
4f density matrix
$\rho_{mm'}^{LDA}$ at the Ce site in the
following way,

\begin{equation}
\Gamma_{mm'}(\varepsilon)=-Im\left\{
\lim_{\eta\rightarrow0}\left[\left(\int
dz\frac{\rho_{mm'}^{LDA}(z)}{\varepsilon-i\eta-z}\right)\right]^{-1}\right\}
.\label{eq:hib}\end{equation}

In all cases the labels \emph{m} and \emph{m'}
correspond to the different
irreducible representations of the 4f states at
the cubic Ce site. That is, for $J=\frac{5}{2}$
the doublet $\Gamma_7$ and the quartet $\Gamma_8$, while
for $J=\frac{7}{2}$ the doublets $\Gamma_6$   and $\Gamma_7$
and the quartet $\Gamma_8$.

The $U\rightarrow\infty$ limit of the Anderson
impurity model is
solved by using the slave boson technique
within the non crossing approximation (NCA).
The NCA equations consist of a couple of
integral nonlinear equations for 
the pseudo-boson ($f^{0}$) and pseudo-fermion
($f^{1}$) self-energies, each of them
containing
the hybridization function
$\Gamma_{mm'}(\varepsilon)$. This $U\rightarrow\infty$ limit is reasonable 
to obtain  crystal field splittings, because the splittings  are stable within 
a wide temperature range above the Kondo temperature. For a detailed
review of the NCA formalism see Ref \cite{Bickers, NCA-bickers-cox}.

With the  LDA-NCA technique one can  calculate 
the crystal-field
splittings and also obtain  the symmetry of the
ground state and excited \emph{4f} levels.
The crystal-field splittings are read from the
separation of the peaks of the 
different spectral functions, $\rho_{mm}$'s,
which are shifted one with respect to the
other due to the different degree of
hybridization of each \emph{4f}
level with the conduction band.
We focus on the value of the splitting corresponding to the 
$J=\frac{5}{2}$ multiplet, namely  
$\Delta_{CF}=\varepsilon_{f\Gamma_7}-\varepsilon_{f\Gamma_8}$.

\section{Results}

The LDA  calculations are performed  at the
experimental
volumes for the CeX and CeY compounds. The
\textit{muffin-tin} radii, $R_{mt}$ are taken
equal to
2.4 a.u. in the case of Ce, while the
corresponding radii
for the anion ligand varies from 1.6 a.u. to 2.8
a.u. depending on
the atomic radius. 102 \textbf{k} points in the
irreducible Brillouin
zone are considered  to be enough  for the
quantities to be calculated.

In the NCA-equations the 4f state has a bare energy value, which we
take  from photoemission experiments.  It is namely,  -3 eV
for the series CeX , -2.6 eV for CeTe and -2.4 eV for, both,
 CeS and CeSe.  All these energy
levels are given with  respect to the Fermi energy \cite{photoemvalues}.

Within NCA the self energies and the Green's
functions are self-consistently obtained by considering the
spectrum up to 6500 K above the Fermi level for the more delocalized
systems (Y=S,Se, X=P,As) and up to 5000 K for the more localized
ones (Y=Te, X=Sb).

\subsection{Crystal-field splittings in
CeX\label{sub:CEF-splittings }}

\begin{figure}[H]
\begin{centering}\includegraphics[clip,scale=0.44]{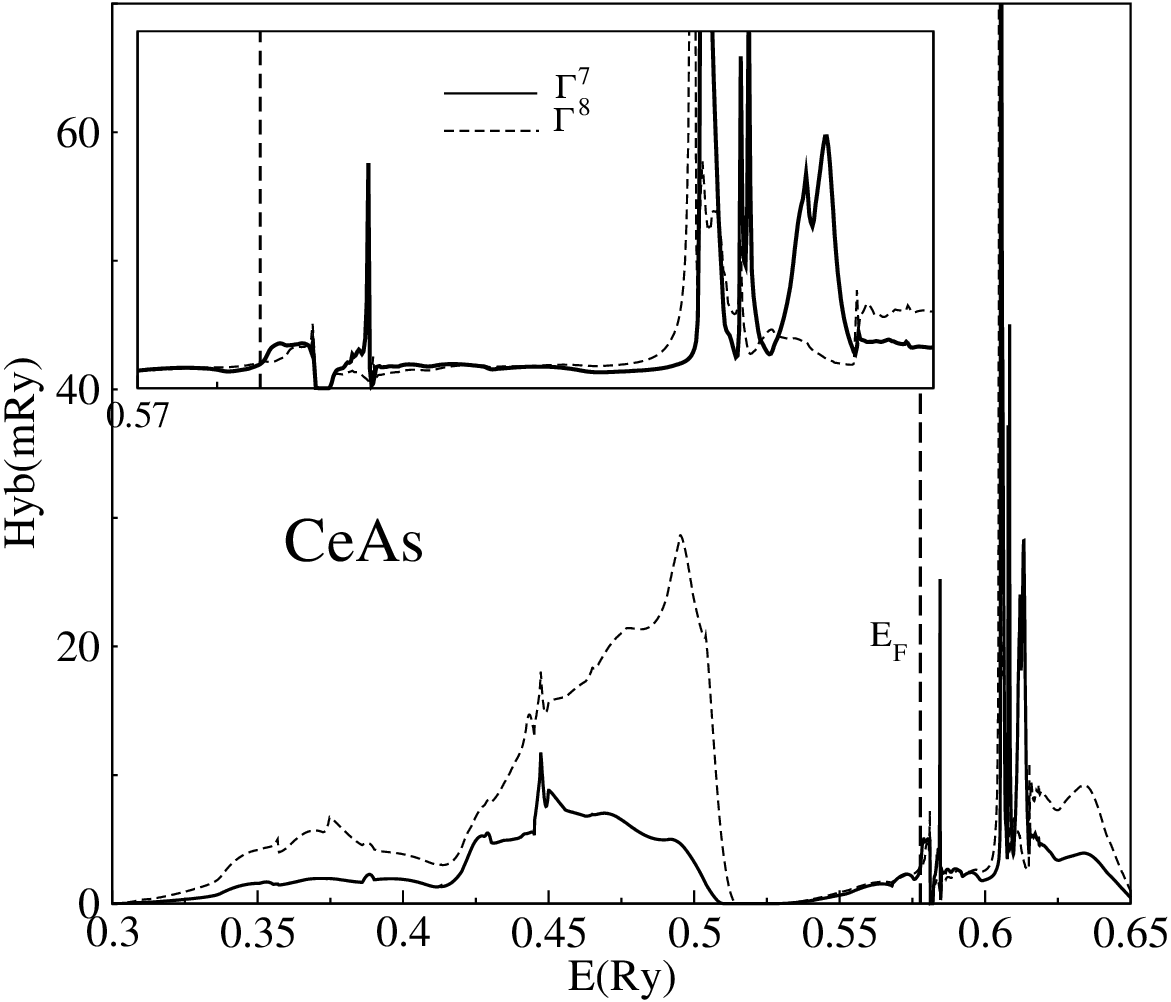}\par\end{centering}
\caption{Hybridization function of the \emph{4f} states with the
conduction band for CeAs : $\Gamma_{7}$
(solid curve) and $\Gamma_{8}$
(dashed curve) symmetries. In the inset, a zoom of an 
energy range around the Fermi level is given.\label{cap:HybAs}}
\end{figure}

We discuss in this section the main features of the 
hybridization functions along the CeX series in order
to understand the evolution of the crystal-field splittings.
As a representative example, we show in 
Figure \ref{cap:HybAs} the calculated
hybridization functions for the $\Gamma_{7}$ and $\Gamma_{8}$ symmetries
corresponding to CeAs.  In the inset, the detailed structure
within an energy window around the Fermi level is given. 

It can be observed that the hybridization functions are 
very rich in structure.  Far below and
far above the Fermi level  the $\Gamma_{8}$ symmetry  is the one 
with the largest hybridization contribution, while close to the Fermi 
level  the $\Gamma_{7}$ one is  stronger. This feature,
with varying relative weights,  holds on
for all the systems under study. The different relative weights
are the finger prints  which  determine the evolution  
of  the crystal-field splittings.  
As it will be shown in the next section, the $4f$ states                                                          
with $\Gamma_{7}$ symmetry
hybridize mainly with the $5d$-Ce
band  while the ones with $\Gamma_{8}$ symmetry 
do it mainly with the anion p states. 

Due to the ionic character of the Ce compounds under study, the LDA unoccupied part of
the spectrum is particularly ill given. Taking this into account, in
in order to reproduce the
trends of the $\Delta_{CF}$'s,  we consider in this work the energy
spectrum  up to an energy of the order of  6000 K above the Fermi level.
Together with the correct  trends  in the splittings,  we also  obtain the experimental 
ground state symmetry, namely the  $\Gamma_{7}$, for all these  systems.  
These results show that for  these compounds the 
energy spectrum around the Fermi energy determines the  symmetry of 
the ground state, even if the $\Gamma_{8}$ hybridization function is the
strongest one in the average. 

The calculated  values for the  splittings
are: 165 K for CeP, 155 K for CeAs, 70 K for CeSb
and 50 K for CeBi.  The obtained values 
and trends  are  in agreement with experiments  and with  the previous
theoretical work.   The calculations reproduce properly 
the sharp decrease of the  splittings when going from CeAs to CeSb.

In order to understand the depression of
$\Delta_{CF}$ in CeSb and
CeBi as compared to CeP and
CeAs\cite{splittingsCeX}, we perform a
calculation
for CeP but at  the  experimental volume of
CeBi, which is the largest one in the series.
The effect of the negative pressure is, as
expected, a reduction in the
strength of the hybridization functions due to
larger interatomic distances. Around the Fermi level this reduction
is  more pronounced for the $\Gamma_{7}$ than for the
$\Gamma_{8}$ symmetry. This decrease in the hybridization  difference between 
both symmetries  gives rise to a  decrease  in the value
of the splitting. That is, the  observed jump in the splittings along 
the series is correlated with a jump in volume when  going 
from  CeAs to CeSb.  This analysis also applies 
to the CeY series (Ce monochalcogenides).\\

\subsection{Crystal field  splittings in CeY vs
CeX\label{sub:CEF-splittings }}

\begin{figure}[H]
\begin{centering}\includegraphics[clip,scale=0.44]{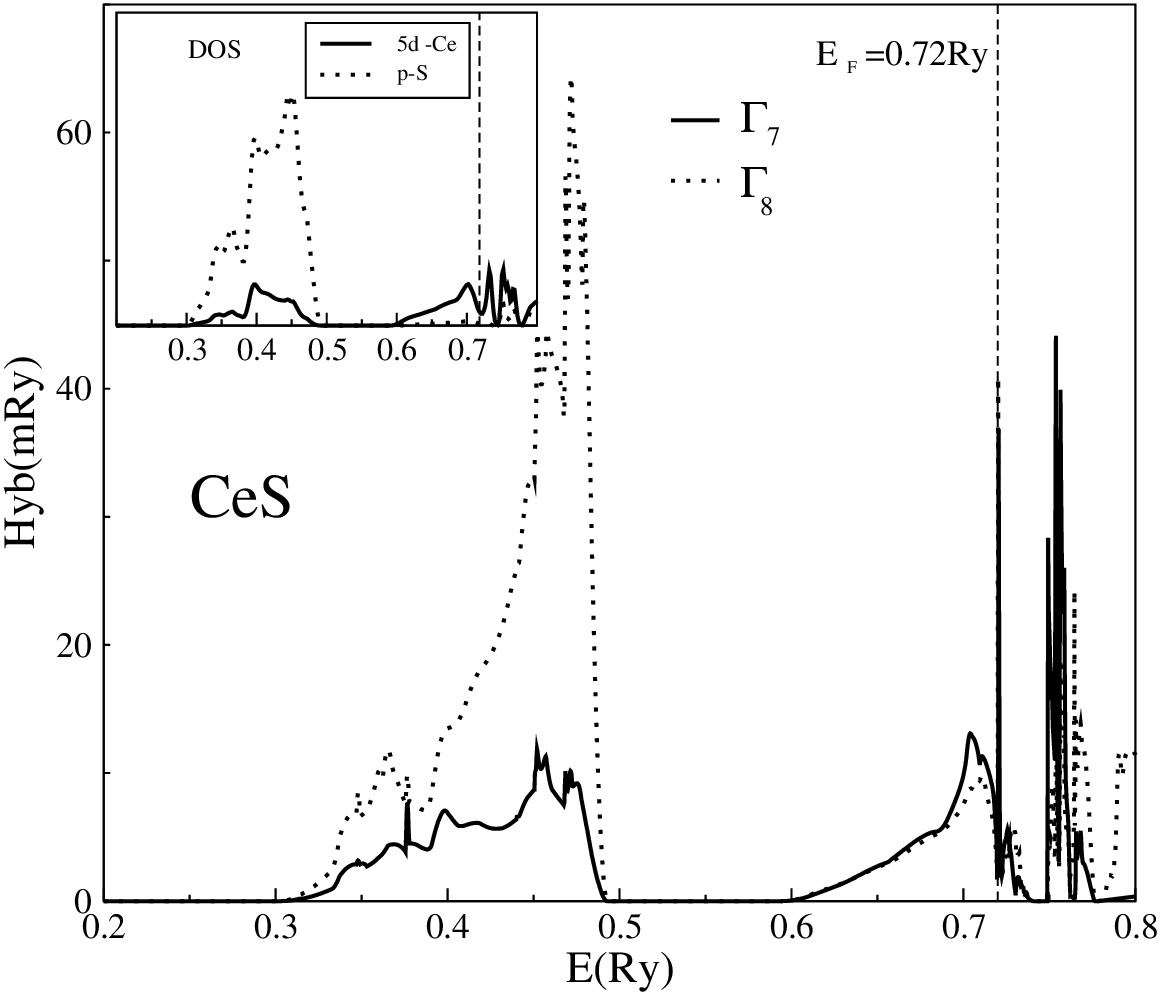}\par\end{centering}

\caption{Calculated hybridization functions for
CeS: $\Gamma_{7}$ symmetry
(solid curve) and $\Gamma_{8}$ symmetry (dotted
curve). Inset: $5d-$DOS
for Ce (solid curve) and $p-$DOS for the anion S
(dotted curve).\label{cap:hybdos-CeS}}
\end{figure}

\begin{figure}[H]
\begin{centering}\includegraphics[clip,scale=0.44]{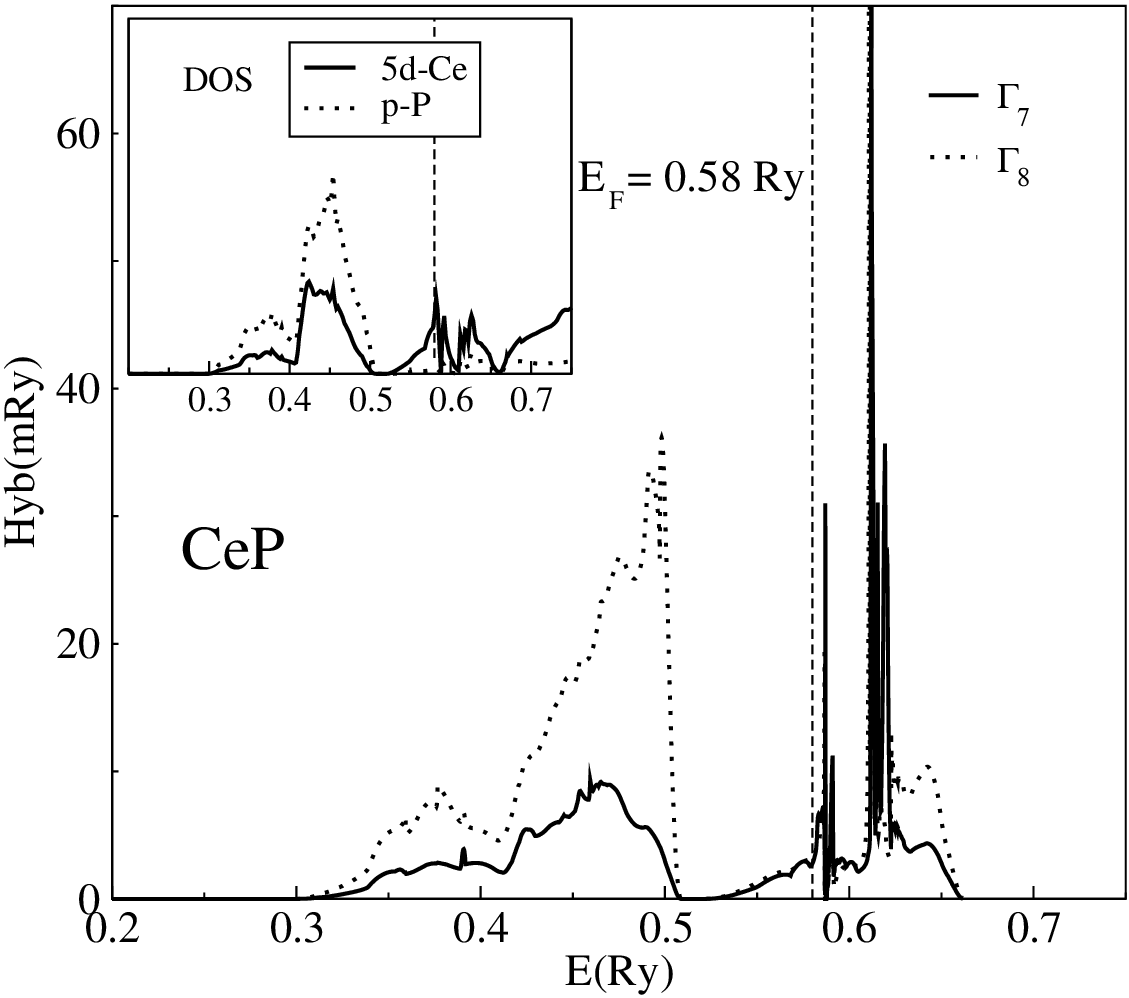}\par\end{centering}

\caption{Calculated hybridization functions for
CeP: $\Gamma_{7}$ symmetry
(solid curve) and $\Gamma_{8}$ symmetry (dotted
curve). Inset: $5d-$DOS
for Ce (solid curve) and $p-$DOS for the anion P
(dotted curve).
\label{cap:hybdos-CeP}}
\end{figure}

Due to the extra p-electron of the Y-anion, the $\Gamma_{8}$ symmetry is more
strongly hybridized in the  CeY series  than
along the CeX one.  The $\Gamma_{7}$ hybridization  is of the
same order of magnitude in both series, as it
can be drawn from the comparison of Fig.
\ref{cap:hybdos-CeS} and Fig.
\ref{cap:hybdos-CeP}. In these figures CeS and CeP are shown as examples.
The same behavior applies for the other compounds.

In the inset of figures \ref{cap:hybdos-CeS} and
\ref{cap:hybdos-CeP}  the partial $p-$ and $5d-$
densities of states are plotted in detail.  Comparing these
densities of states with the corresponding hybridizations,
it can be observed that below $E_{F}$, $\Gamma_{8}$ hybridizes
essentially with the \emph{p}-states
of the X and Y anions, while
$\Gamma_{7}$ does it with
the \emph{5d}-states of the neighbouring Ce atoms. 

\begin{table}[H]
\begin{centering}\begin{tabular}{|c|c|c|c|}
\hline 
Y &
~~ $\Delta_{CF}^{exp}$~~~~$\Delta_{CF}^{NCA}$&
 X &
~~$\Delta_{CF}^{exp}$~~~~$\Delta_{CF}^{NCA}$\tabularnewline
\hline
S &
140~~~~~150&
 P &
172~~~~~165\tabularnewline
\hline
Se&
116~~~~~120&
 As &
159~~~~~155\tabularnewline
\hline
Te &
32~~~~~50&
 Sb &
37~~~~~70\tabularnewline
\hline
\end{tabular}\par\end{centering}

\caption{Experimental and LDA-NCA results for
the crystal-field splittings,
$\Delta_{CF}$=$E(\Gamma_{8})-E(\Gamma_{7})$ , in
the CeX and CeY
series. The splittings are given in Kelvin. See
Ref \cite{splittingsCeY,splittingsCeX} 
for the experimental data.
\label{cap:tabla-cex-cey}}
\end{table}

The values of the calculated $\Delta_{CF}$'s, shown in
Table \ref{cap:tabla-cex-cey} are in very good agreement with the experimental
results. The  order of magnitude as well as the evolution of the 
$\Delta_{CF}$'s  are similar in both series.
It should be noticed that the values  corresponding  
to the CeY  compounds  are slightly smaller than the ones of  
the CeX series,  and this is well reproduced in the calculations. 

In spite of the larger values of the  $\Gamma_{8}$
hybridization function for energies lying far below the
Fermi level,  the splittings are determined mainly
by the hybridization character near the Fermi energy. The extra electron
of the CeY compounds, as compared to the CeX
ones,  does not change   the symmetry of the
ground state  nor that of the  excited ones, preserving the
order of magnitude of the $\Delta_{CF}$'s
when going from CeX and CeY. 

The feature which differentiates the two
series under study, namely, the slightly smaller splitting
observed in the CeY series as compared
to the CeX one, can be correlated to the behavior of
the partial \emph{5d} densities of states at the Fermi level. 
As it can be observed in the \emph{inset}
of Fig. \ref{cap:hybdos-CeS} and
\ref{cap:hybdos-CeP}, the Fermi
level falls in a maximum of the \emph{5d}-DOS
for  CeP and in a valley
of the \emph{5d}-DOS in the case of CeS. 
The $p$-DOS,  on the other hand,  does not change 
considerably  when going from CeP to CeS. 
The larger value of the $5d$-DOS near $E_F$ implies a larger 
$\Gamma_7$ hybridization, which in turn gives rise to a larger 
splitting for CeP than for CeS. This behavior of the partial \emph{5d}-DOS
at the Fermi level also holds for the other  compounds.

\subsection{From CeX to CeY: Crystal-field  splittings in the 
CeX$_{1-x}$Y$_{x}$ alloys}

\begin{figure}
\begin{centering}\includegraphics[clip,scale=0.35]{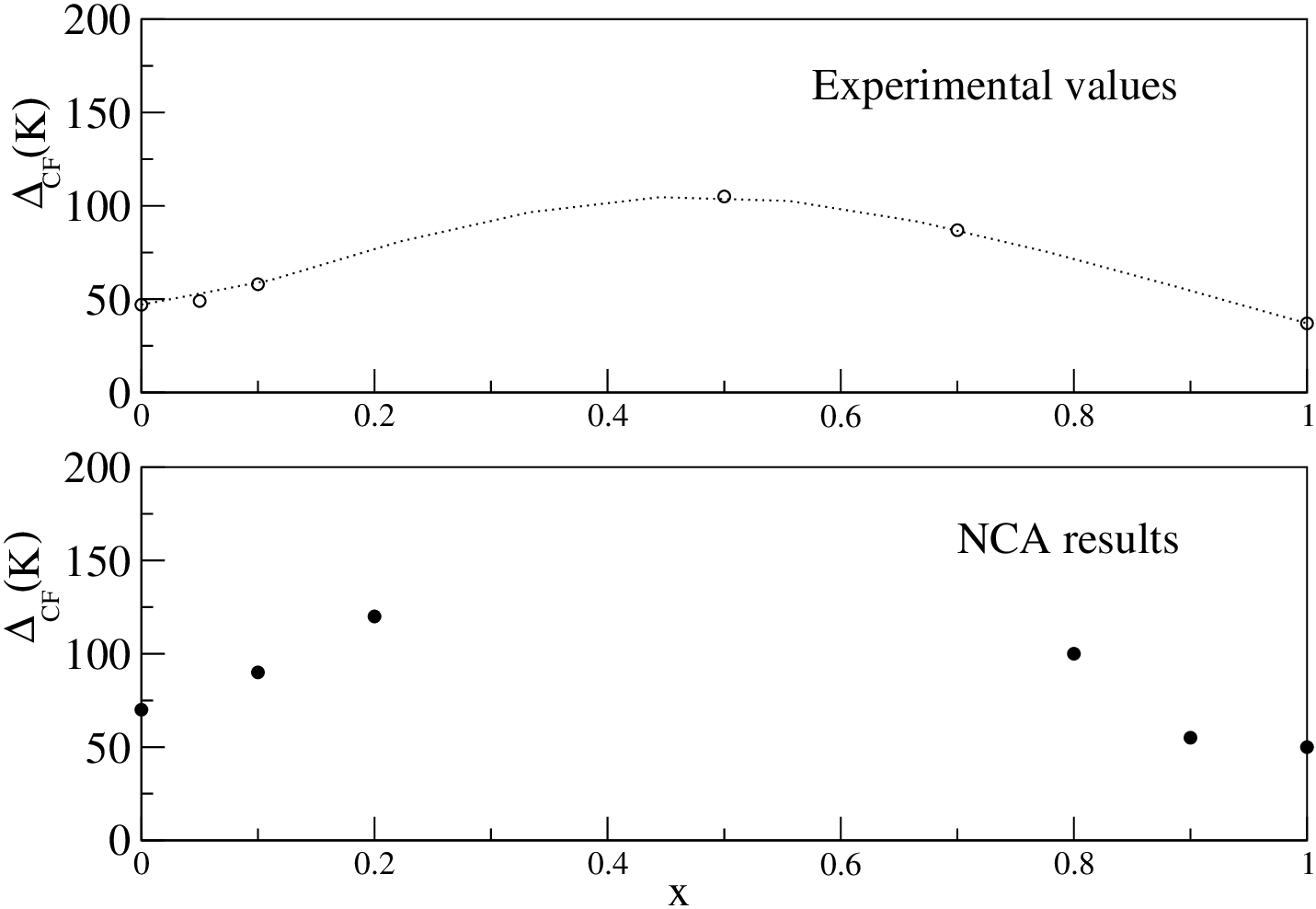}\par\end{centering}

\caption{Calculated $\Delta_{CF}$'s of
CeSb$_{1-x}$Te$_{x}$ systems. We have
denoted CeSb+0.1 and CeSb+0.2 as $x=0.1$ and
$x=0.2$ and CeTe-0.2
and CeTe-0.1 as $x=0.8$ and $x=0.9$,
respectively. Experimental
and calculated results are given in Kelvin. See
Ref. \cite{CeXYalloys} for experimental
data. \label{fig:CeSb1-xTex}}
\end{figure}

So far, until now we have made a comparative analysis of the
evolution of the $\Delta_{CF}$'s along the two series CeX and CeY, the
difference between them being the extra p-electron of the Y atoms
as compared to the X ones. The $\Delta_{CF}$'s 
of the  solid solutions CeSb$_{1-x}$Te$_{x}$
were measured  by inelastic neutron scattering by \emph{Rossat} 
\cite{CeXYalloys} for different \emph{Te}-concentrations
( x= 0.05, 0.1, 0.5 and 0.7).
The experimental data  are  shown in the upper plot of
Figure \ref{fig:CeSb1-xTex}. To the best of our knowledge, there
is no previous theoretical work interpreting the shown  results and 
complementing the suggestions done by Rossat in the sense that 
the observed trends in the alloys should be associated to 
$p-f$ mixing.

In order to understand  the non linear evolution of
$\Delta_{CF}$ as a function of Te concentration, we
simulate the substitution of Sb by Te, and viceversa, by doing 
electronic structure calculations  within  the virtual-crystal 
approximation\cite{VCA1,VCA2} (VCA) in the two concentration
limits.  We have also analysed the splittings of the lighter 
CeAs$_{1-x}$Se$_{x}$ alloys
within the same approximation. To this  purpose, we add  a small  
amount of valence  electrons  to the  X atom of CeX  and  withdraw 
small  amounts of valence electrons 
from the Y atom of CeY.  Charge neutrality is preserved in the 
calculations.
Nearly one third of the added charge goes to the \emph{p} levels 
of the anions, while the rest goes to the  interstitial region.
The occupation of the \emph{5d} and \emph{4f} levels of the Ce atoms,
remains unchanged. Within this approach  the original crystal symmetry is
preserved.

In the case of CeSb$_{1-x}$Te$_{x}$, we calculate the values of the 
splittings for  four different VCA concentrations, 
namely by adding 0.1- and  0.2-electrons 
to  \emph{Sb}  and by subtracting  0.1- and 0.2-electrons from   \emph{Te}
in  the  parent compounds \emph{CeSb} and \emph{CeTe}, respectively. In order
to compare with the behavior in the lighter alloys CeAs$_{1-x}$Se$_{x}$,
we consider  for these last systems  two concentrations, 
one adding 0.2e to As and the other substracting this same amount of electrons 
from Se.  In all cases the VCA calculations are done at the experimental
volume of the corresponding parent compounds, as we are just 
studying  here  the  electronic contribution to the evolution of 
the $\Delta_{CF}$ values.  In the first case, the obtained splittings 
are $\Delta_{CF}(CeSb+0.1e)=90$ K,
$\Delta_{CF}(CeSb+0.2e)=120$ K, $\Delta_{CF}(CeTe-0.1e)=55$ K and
$\Delta_{CF}(CeTe-0.2e)=100$ K. For the second simulated alloys the values are
$\Delta_{CF}(CeAs+0.2e)=200$ K and $\Delta_{CF}(CeSe-0.2e)=200$ K. 
According  to  the experimental trends for CeSb$_{1-x}$Te$_{x}$,
we obtain an increase  in the value of the crystal-field splittings
when a small amount of charge is added to CeSb and also  when
it is  taken away from CeTe, as it is shown in Figure \ref{fig:CeSb1-xTex}.
For the lighter  alloys there is no available experimental results for 
the splittings, but we obtain the same behavior as for the other ones. 

\begin{figure}
\begin{centering}\includegraphics[clip,scale=0.35]{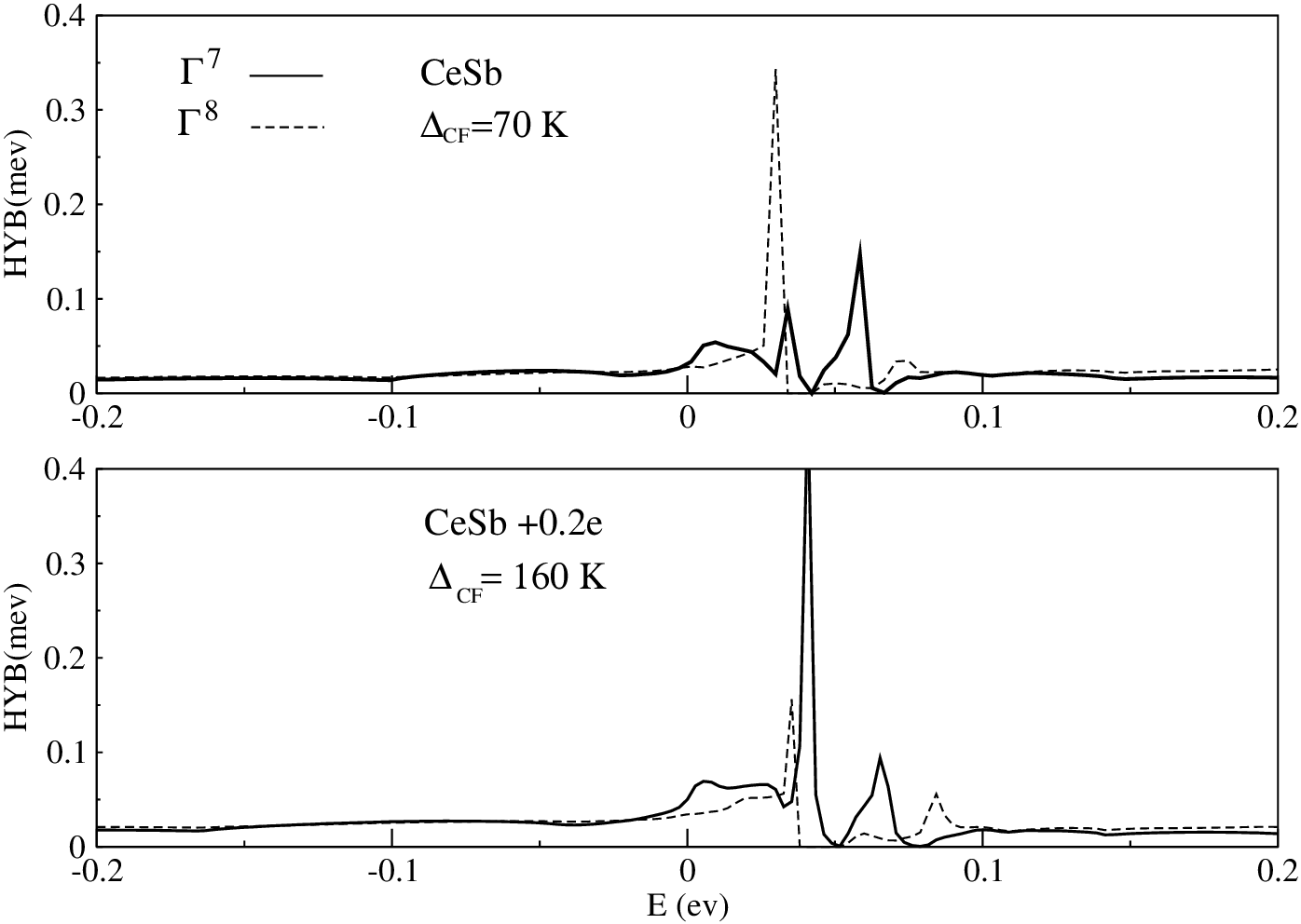}\par\end{centering}

\caption{$\Gamma_{7}$and $\Gamma_{8}$
hybridization functions around the
Fermi level for: CeSb (\emph{upper
pannel}) and CeSb+0.2 (\emph{lower
pannel}).\label{fig:cesb+0.2-hyb}}
\end{figure}

When \emph{CeX} is doped with a  small amount
of charge the value of the $\Gamma_{8}$ hybridization function
increases slightly for energies  below $E_{F}$. This is 
due to an increasing \emph{4f}-\emph{p} mixing in
that energy region. This extra charge also affects indirectly the 
\emph{4f}-\emph{5d} 
hybridization  through an increased \emph{p}-\emph{5d} mixing.
This indirect mechanism induces a slight increment of the $\Gamma_{7}$ 
hybridization function around $E_{F}$.  
There is no considerable contribution to the
crystal-field
splittings  from energies lying more than 0.03 Ryd
below $E_{F}$,  because the hybridization
functions
of both symmetries change by a similar amount.
This can
be seen in Figure \ref{fig:cesb+0.2-hyb}.
However, above $E_{F}$,
the $\Gamma_{8}$ hybridization goes down due to
a decreasing number
of unoccupied \emph{p} states, while the
$\Gamma_{7}$ hybridization function
goes up, these two effects  give rise to an
increase in the value of $\Delta_{CF}$.
These results agree with the experimental ones 
obtained by Rossat \emph{et al.} 
and  the output of our calculations reinforce
the interpretation done by them.

\begin{figure}
\begin{centering}\includegraphics[clip,scale=0.35]{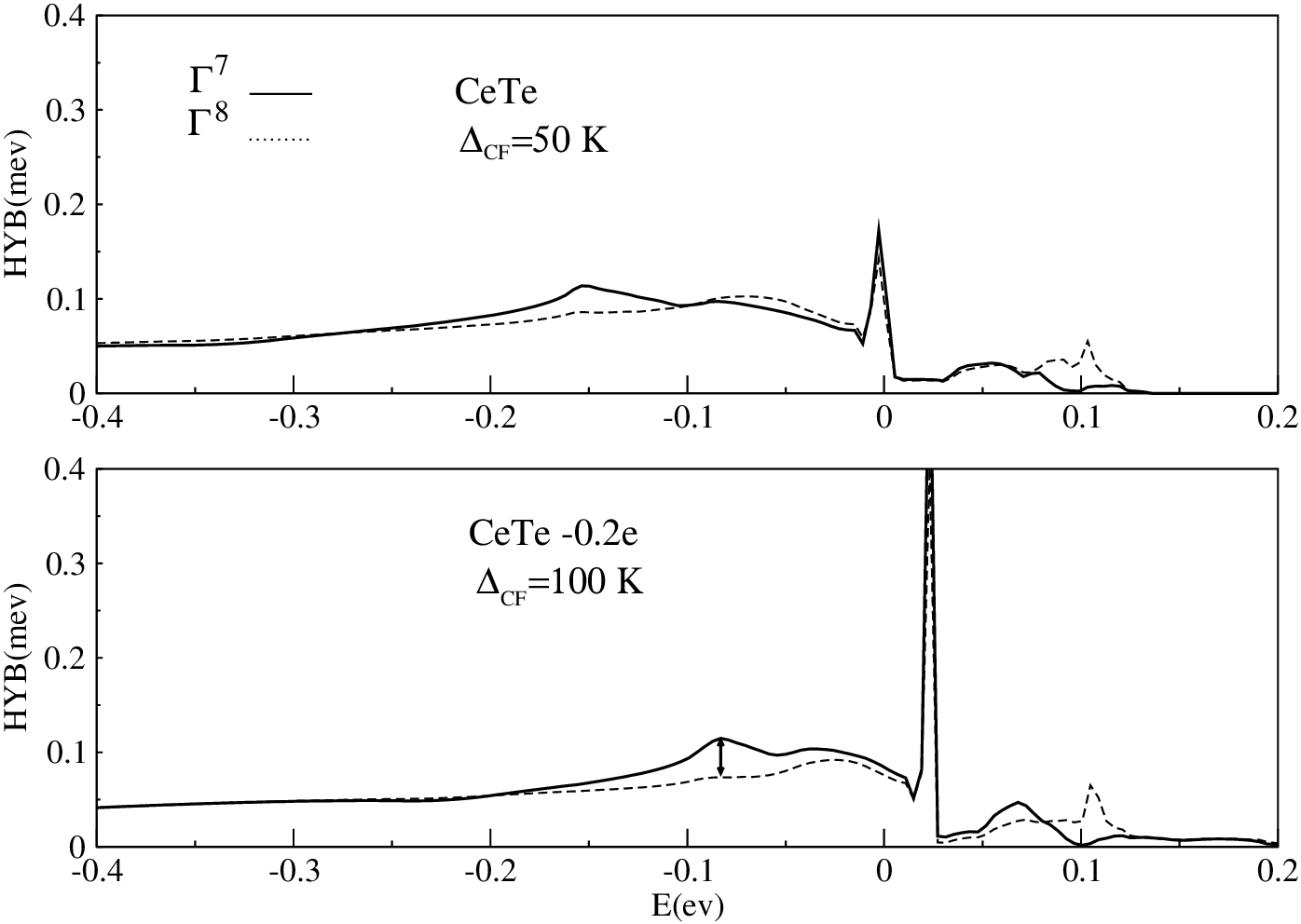}\par\end{centering}

\caption{$\Gamma_{7}$and $\Gamma_{8}$
hybridization functions around the
Fermi level for: CeTe (\emph{upper
pannel}) and CeTe-0.2 (\emph{lower
pannel}).\label{fig:CeTe-0.2hyb}}
\end{figure}

In the other concentration limit, when a small
amount of charge is removed
from Te in CeTe, there is, in the
average,  a slight decrease of
the hybridization strengths below the Fermi
level for  both symmetries ,
namely  $\Gamma_{7}$ and $\Gamma_{8}$, but
$\Gamma_{8}$ is the one which shows the
largest decrease. This last effect is to be
attributed  
to the fact that $\Gamma_{8}$  comes mainly from
\emph{p}-\emph{4f} mixing and, that 
the withdrawn charge is essentially of \emph{p}
character.   
This growing difference between both
hybridization strengths is the reason for 
the experimentally observed larger values of the  crystal
field splittings, as compared to the splitting
corresponding to CeTe.
In Figure \ref{fig:CeTe-0.2hyb} it can be seen
that
$\Gamma_{7}$ is larger than $\Gamma_{8}$ 
below $E_{F}$ for the system $CeTe-0.2e$ (lower
pannel). In this plot,
we indicate the symmetry separation (between
$\Gamma_{7}$ and $\Gamma_{8}$) with an arrow.

\section{Discussion and conclusions}

We apply in this work  an \emph{ab initio}-many body technique (LDA-NCA)  to calculate
the crystal-field splittings of the monopnictides CeX and
the monochalcogenides CeY series and their alloys. Within the 
$ab initio$ frame, in which the 4f states are treated as part of the
valence band,  we calculate the hybridization function 
$\Gamma_{mm'}(\varepsilon)$, which contains detailed information
on the electronic structure of each material.  This $\Gamma_{mm'}(\varepsilon)$
function is used as input for the Anderson
impurity model hamiltonian,  which is solved
within a multi orbital non-crossing
approximation in the $U\rightarrow\infty$ limit. 

We show that the LDA-NCA technique  gives the correct trends
for the evolution of the crystal-field  splittings of these systems,  
being the results in good agreement with the available experimental data \cite{splittingsCeX} and with 
previous theoretical calculations \cite{point-charge-1}.
We also obtain  the correct symmetry for the ground  state of the 4f multiplet, that is, the 
$\Gamma_{7}$ one for all the compounds and alloys.
In particular, we correlate  the sudden decrease of the $\Delta_{CF}$ when going from CeAs to CeSb 
(which are isoelectronic), to a big volume change.

The most important contribution to the values of the crystal-field splittings comes from energies around 
the Fermi level. We obtain that the hybridization strength around $E_{F}$ is due mainly 
to the 5d-4f mixing. In those systems where the 4f states are more localized (CeBi, CeSb, CeTe) the
intensity of the 5d-4f and of the p-4f hybridizations is smaller than in those where they  are more 
delocalized. Due to this facts,  magnitudes that depend on them, such as the crystal field splittings,
are atenuated with respect to those compounds in which the 4f states  more delocalized.

When comparing the results obtained for CeX with those for CeY, 
we obtain that the outcoming crystal-field splittings of the first series are 
slightly larger than those of the second  one even if the CeY series has one more p
 electron. This is attributed to the fact that in the CeX series the Fermi 
level falls in a maximum of the 5d-DOS while in the CeY one it 
does it in a valley.  The effect of the extra p electron in CeY is to produce
an enlargement of  the $\Gamma_8$ hybridization far below the Fermi level, and this  does 
not  influence the value of the splitting considerably. 
 
Finally, we  analyse the evolution of the splittings in the CeX$_{1-x}$Y$_{x}$ alloys
as a function of x. We consider one of the 'heavy' alloys, namely CeSb$_{1-x}$Te$_{x}$, 
and one of the  'light' ones, CeAs$_{1-x}$Se$_{x}$.  Alloying gives rise to a non monotonic 
evolution of the $\Delta_{CF}$'s which according to our results, obtained within the 
virtual-crystal  approximation,  can be explained as an electronic effect by the only
consideration of the added extra charge and to its effect on the hybridization functions.
\\

\section{ACKNOWLEDGMENTS}

This work was partially funded by UBACyT-X115,
PICT-0310698 and PIP
2005-2006 Num. 6016. A. M. Llois and V. L.
Vildosola belong to CONICET (Argentina).

\bibliographystyle{apsrev}
\bibliography{../tesis.bib}

\end{document}